\documentstyle[aps,multicol,epsf,subfigure]{revtex}
\def\switchmulticol#1{#1}

\def\D{\partial}
\def\grad{\nabla}

\def\lap{\nabla^2}

\def\etal{{\it et al.}}

\def\Ref#1{(\ref{#1})}
\def\Eq#1{Eq.(\ref{#1})}

\def\eq{\begin{eqnarray}}
\def\qe{\end{eqnarray}}
\def\eqnn{\begin{eqnarray*}}
\def\qenn{\end{eqnarray*}}
\def\nn{\nonumber}

\def\bq{\bm{q}}
\def\br{\bm{r}}
\def\bu{\bm{u}}
\def\bv{\bm{v}}

\def\simge{\;\lower3pt\hbox{$\stackrel{\textstyle >}{\sim}$}\;}
\def\simle{\;\lower3pt\hbox{$\stackrel{\textstyle <}{\sim}$}\;}
\def\bm#1{\mbox{\bf #1}}

\def\lrL#1{\left[#1\right]}

\def\lrS#1{\left(#1\right)}
\def\lrF#1{\left|#1\right|}

\def\biglrL#1{\bigl[ #1 \bigr]}

\def\BiglrL#1{\Bigl[ #1 \Bigr]}

\def\BiglrF#1{\Bigl| #1 \Bigr|}

\def\bigglrL#1{\biggl[ #1 \biggr]}
\def\bigglrM#1{\biggl\{#1 \biggr\}}
\def\bigglrS#1{\biggl( #1 \biggr)}

\def\mycomment#1{}
\def\mycaption#1{\nopagebreak[4]\hspace{0mm}\\\begin{minipage}[h]{85mm}\caption{#1}\end{minipage}\vfill}

\def\f#1#2{\frac{#1}{#2}}

\newcommand{\bqp}{{\bf q}_\perp}
\newcommand{\bqtp}{\mbox{\scriptsize\bf q}_\perp}
\renewcommand{\bu}{\mbox{\bf u}}
\renewcommand{\bv}{\mbox{\bf v}}
\newcommand{\but}{\mbox{\scriptsize\bf u}}
\newcommand{\bvt}{\mbox{\scriptsize\bf v}}
\newcommand{\cH}{{\cal H}}
\newcommand{\cI}{{\cal I}}
\newcommand{\cJ}{{\cal J}}

\newcommand{\Zhat}{\widehat{Z}}

\title  {Casimir Effect in Fluids above the Isotropic-Lamellar Transition}
\author {Nariya Uchida %
}
\address{Department of Physics, Kyoto University, Kyoto 606, Japan}
\date{%
Submitted 2 Sep 2000; revised 14 Dec 2000; published 31 Oct 2001
}

\begin{document}
\draft

\bibliographystyle{prsty}

\maketitle
\begin{abstract}
We study fluctuation-induced interaction
in confined fluids above the isotropic-lamellar transition.
At an ideally continuous transition,
the disjoining pressure has the asymptotic
form $\Pi(d\to\infty)\approx -C k_BT q_0^2/d$,
where $d$ is the interwall distance, 
$q_0$ is the wavenumber of the scattering peak,
and $C= 1/(4\pi)$ in the strong anchoring limit. 
The long-rangedness is enhanced due to continuous 
distribution of soft modes 
in the ${\bf q}$-space.  An unconventionally 
strong Casimir force with a range of several 
lamella thicknesses is realistic above the transition. 
We also find an oscillatory force profile near 
a surface-induced transition.
\end{abstract}


% 68.15.+e Liquid thin films
% 64.60.Cn Order-disorder transformations; statistical mechanics of
%          model systems
% 61.25.Hq Macromolecular and polymer solutions; polymer melts; swelling
% 82.70.-y Disperse system

\switchmulticol{\begin{multicols}{2}}

Fluids in confined geometries show
a variety of phenomena that are not observed in bulk.
Among them, thermal-fluctuation--induced
interactions between boundary walls
or objects immersed in the fluid
have attracted much attention. 
They are called the Casimir forces by analogy 
with the quantum original~\cite{Casimir}.
The forces are long-ranged in systems with 
soft modes, such as critical simple 
fluids~\cite{FdG,Krech} and liquid crystals~\cite{Mikheev,APP},
with the decay law and amplitude modulated by 
various surface effects~\cite{Krech,LK,ZPZ}.
Recent evidence~\cite{ML,GC,VVC} supports the view 
that the interaction is ubiquitous in correlated fluids~\cite{KG}.

In this Letter, we address the effect 
in structured fluids above the isotropic-lamellar (I-L)
transition. The model system we consider has 
two characteristic lengths, which describe decay 
and oscillation of the correlation 
function. 
Physical realizations of the model
include block copolymers in the disordered phase
and bicontinuous microemulsions. 
Films of block copolymers have constituted 
the subject of numerous papers~\cite{Turner,SSKBSM,KWMLRGS,MM,BFS}.
Most of them focus on the lamellar phase,
while a few theoretical works treat
the system above the I-L
transition~\cite{MM,BFS}.
The latter works discuss the effect of a surface field,
which induces a mean-field interaction
between boundary walls.
In comparison to the case of copolymers, much less 
is known about confined fluids containing 
short-chain surfactants~\cite{SS,HO,CTM}.
A mean-field study with a standard Ginzburg-Landau
model of microemulsions found a surface-induced 
I-L transition that preempts the bulk transition~\cite{SS}.

The bulk property of the system we assume is 
described by the Hamiltonian
\eq
\cH_b = \f{\epsilon}{2} \int \f{d\bq}{(2\pi)^3} \bigglrL{(q^2-q_0^2)^2 + p_0^4} 
\lrF{\phi_{\bq}}^2,
\label{Hfourier}
\qe
where $\epsilon$ is a constant,
$\phi$ is the order parameter,
and $p_0$ and $q_0$ are characteristic wavenumbers
that generally depend on the temperature.
This and similar types of Hamiltonians,
with or without additional nonlinear terms,
have been applied to various kinds of complex
fluids~\cite{Leibler,TS,ABJ,Swift}.
For symmetric AB-block copolymer melts, for instance, 
$\phi$ represents
the excess of A-monomer's (say) volume fraction
with respect to its spatial average, and
the parameters are given by
$\epsilon=0.0327 k_BT R_g /N^{1/2}$, $q_0=1.95/R_g$,
and $p_0=1.43 [(\chi_c - \chi) N]^{1/4}/R_g$.
Here,
$R_g$ is the chain's gyration radius,
$N$ is the polymerization index,
$\chi$ is the Flory interaction parameter, and 
$\chi_c=10.5/N$ locates the mean-field I-L transition~\cite{Leibler}. 
To be general, we shall use the ratio $p_0/q_0$ as the measure of 
distance from the transition temperature.

We consider a confined film of thickness $d$ and 
of macroscopic area.
The total Hamiltonian is the sum of 
bulk and surface contributions,
${\cal H} = {\cal H}_b + {\cal H}_s$.
We write the bulk Hamiltonian \Ref{Hfourier} 
in the real space as
\eq
{\cal H}_b &=& \f{\epsilon}{2} \int_{0<z<d} d\br
\bigglrL{(\lap \phi)^2 - 2 q_0^2 (\grad \phi)^2 + (p_0^4 + q_0^4) \phi^2
}.
\label{hamiltonian}
\qe
In this form, the Hamiltonian is equivalent to
the Teubner-Strey model of microemulsions~\cite{TS}
in the temperature region between the bicontinuous-lamellar transition 
and the Lifshitz point.
For chemically neutral and identical walls, 
the surface part of the Hamiltonian 
can be expanded up to the second order in $\phi$ and $\grad \phi$
as~\cite{GZ}
\eq
{\cal H}_s= \int d\br 
\lrL{\omega_s \phi^2 + g_s (\grad\phi)^2} 
\biglrL{\delta(z)+\delta(z-d)}.
\label{Hs}
\qe

The free energy of an elastic medium in a 
finite geometry
can be computed by several different
techniques for regularizing a divergent
sum~\cite{APP,LK,ZPZ,Fierz}.
The analysis reduces to a 1D problem
for the in-plane Fourier modes 
$\phi_{\bqtp}(z) = \int d\br_\perp \exp(-i\bqp\br_\perp) \phi(\br)$,
where $\br_\perp=(x,y)$ and $\bqp=(q_x,q_y)$.
The partition function for each in-plane mode 
is given by
\eq
Z_{\bqtp} &=& \int\!d\bu\!\int\!d\bv\! 
\int_{\but,\bvt}
\hspace{-3mm}
{\cal D}\phi_{\bqtp} 
e^{-
\lrL{ 
{\cal H}_{b,\bqtp} + (\omega_s + g_s q_\perp^2)\but^2 + g_s \bvt^2
}/k_BT}.
\label{Zcond}
\qe
Here, $\cH_{b,\bqtp}$ is the Fourier component of $\cH_b$
and $\int_{\but,\bvt} {\cal D}\phi_{\bqtp}$ means that 
the functional integral should be taken over 
the paths that satisfy
\eq
\lrS{\phi_{\bqtp}(0), \phi_{\bqtp}(d)} = \bu
\label{BC1}
\qe
and
\eq
\lrS{\phi'_{\bqtp}(0), \phi'_{\bqtp}(d)} = \bv,
\label{BC2}
\qe
where ${}^\prime = \D/\D z$.
The path integral 
for a quadratic Hamiltonian
that contains a squared second-derivative
has been calculated by Kleinert~\cite{Kleinert}.
Following his work,
we decompose each fluctuation path that satisfies
the conditions \Ref{BC1} and \Ref{BC2}
into two parts as
$\phi_{\bqtp}=\phi_{m,\bqtp}+\delta\phi_{\bqtp}$,
where $\phi_{m,\bqtp} = \phi_{m,\bqtp}(z;\bu,\bv)$
is defined as the path that minimizes ${\cal H}_{b,\bqtp}$
under the same boundary conditions.
Then the Hamiltonian is decomposed as
\eq
\cH_{b,\bqtp} &=& \cI_{\bqtp} + \cJ_{\bqtp},
\\
\cI_{\bqtp} &=&  \f{\epsilon}{2}
\int_0^d dz \; 
\bigglrM{
\lrF{
\delta \phi''_{\bqtp}}^2
+ 2 (q_\perp^2 - q_0^2) \lrF{\delta \phi'_{\bqtp}}^2
\switchmulticol{\nn\\&& \hspace{18mm}}
+ \BiglrL{(q_\perp^2 - q_0^2)^2 + p_0^4}
\BiglrF{\delta \phi_{\bqtp}}^2
},
\\
\cJ_{\bqtp} &=&
\f{\epsilon}{2}
\bigglrL{ 
\phi'_{m,\bqtp} \phi''_{m,-\bqtp} 
- \phi_{m,\bqtp} \phi'''_{m,-\bqtp} 
\switchmulticol{\nn\\ && \hspace{20mm}}
+ 2 (q_\perp^2 - q_0^2) \phi_{m,\bqtp} \phi'_{m,-\bqtp}
}^{d}_{0}.
\label{Hdecompose1}
\qe
The latter part
can be expressed in terms of
the boundary values as
$\cJ_{\bqtp} = {\bm a} \cdot {\sf M} \cdot {\bm a}$,
where ${\bm a} = (u_1, u_2, v_1, v_2)$
and ${\sf M}={\sf M}(q_\perp, d)$ is a $4 \times 4$ matrix
whose calculation is straightforward.
Accordingly, the partition function 
is factorized
into ``bulk'' and ``surface'' parts
as
\eq
Z_{\bqtp} &=& Z_{b,\bqtp}\cdot Z_{s,\bqtp},
\\
Z_{b,\bqtp} &=& 
\int_{{\bf 0},{\bf 0}}{\cal D}\delta\phi_{\bqtp}e^{-\cI_{\bqtp}/k_BT},
\\
Z_{s,\bqtp} &=&  
\lrS{
\det \; \f{
{\sf M} + (\omega_s + g_s q_\perp^2) {\sf E}_{\bu} + g_s {\sf E}_{\bv}
}{\pi k_BT}
}^{-1/2},
\\
{\sf E}_{\bu} &=& {\rm diag}(1,1,0,0), 
\\
{\sf E}_{\bv} &=& {\rm diag}(0,0,1,1). 
\qe
The bulk factor can be computed using Lagrange multipliers
that impose the boundary 
conditions $\delta \phi_{\bqtp} = \delta\phi_{\bqtp}'=0$~\cite{Kleinert}.
In order to extract the interaction part
of the free energy, we should regularize
the partition function as
$\Zhat_{\bqtp}(d) = \lim_{D\to\infty}
[Z_{\bqtp}(d)Z_{\bqtp}(D-d)/Z_{\bqtp}(D/2)^2]$,
so that $\Zhat_{\bqtp}(\infty)=1$~\cite{Fierz}.
Applying this to $Z_{b,\bqtp}$,
we have
\eq
\Zhat_{b,\bqtp}(d) &=& 
\f{e^{k_+ d}}{2} \lrL{\sinh^2 (k_+ d) 
- \f{k_+^2}{k_-^2} \sin^2 (k_- d)}^{-1/2},
\label{Vfl1}
\\
k_\pm &=& k_\pm(q_\perp) 
\nn\\&=& 
\sqrt{
\f{1}{2} 
\bigglrL{
\sqrt{(q_\perp^2-q_0^2)^2+p_0^4} \pm (q_\perp^2 - q_0^2)}
}.
\label{Vfl3}
\qe
In the same way, we obtain the 
regularized surface partition function $\Zhat_{s,\bqtp}$,
which has a voluminous expression that cannot 
be written down here.
The interaction free energy per area is obtained as
\eq
F &=& - k_BT \int \f{d \bq_\perp}{(2\pi)^2}
\bigglrS{
\ln \Zhat_{b,\bqtp} + \ln \Zhat_{s,\bqtp}
}.
\label{freeenergy}
\qe

%%%%%%%%%%%%%%%%%%%
% Force profile.
%%%%%%%%%%%%%%%%%%%

First we focus on the bulk contribution by 
taking the ``strong anchoring'' limit $\omega_s\to \infty$, 
$g_s\to\infty$, for which $\Zhat_{s,\bqtp}=1$.
We show the result in terms of 
the disjoining pressure $\Pi= - \D F/\D d$.
At the critical point $p_0=0$,
the contributions to the pressure
from the regions $q_\perp<q_0$
and $q_\perp>q_0$ are given by, respectively,
\eq
\Pi_< &=& -\f{k_BT}{2\pi d^3} \int_0^{q_0 d} ds \;
\f{s^2 (s - \sin s \cos s)}{s^2 - \sin^2 s}
\label{Pi<}
\\&&
\left\{
\begin{array}{ll}
\displaystyle
\approx -\f{k_BT q_0^2}{2\pi d}, & \qquad q_0 d \ll 1,
\vspace{2mm}
\\
\displaystyle
\approx -\f{k_BT q_0^2}{4\pi d}, & \qquad q_0 d \gg 1,
\end{array}
\right.
\qe
where $s$ is for $\sqrt{q_0^2-q_\perp^2}d$,
and
\eq
\Pi_> &=& -\f{k_BT}{2\pi d^3} \int_0^{\infty} dt \;
\f{t^2 (e^{-t} \sinh t + t^2 - t)}{\sinh^2 t - t^2}
\nn\\
&=& -0.5463\ldots \times \f{k_BT}{d^3}, 
\label{Pi>}
\qe
where $t$ is for $\sqrt{q_\perp^2-q_0^2}d$.
Note that the total force $\Pi_< + \Pi_>$
is proportional to $1/d$ at large distances.
To my knowledge, 
it is more long-ranged than any 
(thermal or quantum)
Casimir interaction ever predicted for a 3D system.
As seen from \Eq{Pi<},
the origin of the $1/d$ tail lies in that
every in-plane mode with a wavenumber $q_\perp < q_0$ has a
long-range correlation.
It is further reduced to the fact that
the bulk soft modes of the system 
are distributed over the surface of the
sphere $|\bq| = q_0$. This is in contrast to
the case of a critical simple fluid,
in which the soft mode is located on 
the single point $\bq={\bf 0}$
and the Casimir force is proportional to 
$d^{-3}$ (in the harmonic approximation). 
The extra factor $d^2$ in the present system 
is thus explained by the dimensionality of 
distribution of soft modes in the $\bq$-space.
Shown in Fig.1(a) is the force profile
at the critical point.

%%%%%%%%%%%%%%%%%%%%%%
% Finite Temperature
%%%%%%%%%%%%%%%%%%%%%%
%
In real systems,
nonlinearity in the free energy
makes the I-L transition weakly
first order~\cite{Brazovskii},
and hence we do not expect to see the truly long-ranged force.
For block copolymers, for instance, we estimate the typical and effective
value of $p_0/q_0$ at the transition to be $0.4$.
Shown in Fig.1(b) is the profile of the scaled pressure $\Pi d^3/k_BT$
at finite values of $p_0/q_0$.
Even considerably above the transition,
the force has a range of
several times $\pi/q_0$, the lamella thickness.
A weak oscillation 
arises from the region $q_\perp < q_0$, while
the contribution from the region
$q_\perp > q_0$ is always monotonic.
Note also the large magnitude of the interaction;
the scaled pressure in its decay range is
typically $10$ times larger than the value $\Pi d^3/k_BT=-0.096$
for nematic liquid crystals in the one-constant 
approximation~\cite{APP}.

%%%%%%%%%%%%%%%%%%%%%
% Surface factor.
%%%%%%%%%%%%%%%%%%%%%

Next we study the role of fluctuations at
boundaries. For the present model
with a vanishing surface potential ($\omega_s=g_s=0$),
it is known that the mean-field solution $\phi=0$
is destabilized above the bulk I-L transition~\cite{SS}.
The instability remains for a small but
finite surface potential.
I calculated the region of surface parameters
for which the homogeneous mean-field solution is
stable at any temperature above the bulk I-L transition.
It is plotted in Fig.2(a) in terms of
the dimensionless parameters
$\Omega_s = \omega_s /(\epsilon q_0^3)$
and $G_s = g_s /(\epsilon q_0)$.
Shown in Fig.2(b) is the disjoining pressure
for different values of $(\Omega_s, G_s)$
in the stable region
and at the bulk transition temperature.
The force has
the asymptotic behavior
$\lim_{d\to\infty} \Pi d/(k_BT q_0^2) = -C$,
where $C$ is a positive number that
weakly depends on $(\Omega_s, G_s)$.
An oscillation in $\Pi(d)$ appears
for a small surface potential,
and is amplified as we approach
the instability line.
On the line, the quantity $\Pi d/(k_BTq_0^2)$ is 
almost periodic in $d$ except at small distances.
The peak positions are approximately
given by $d=(n\pi+\psi)/q_0$,
where $n$ is any integer and
the phase shift $\psi$ moves over the region $[0,\pi]$
as we move over the instability line.
To see the origin of the oscillation,
we consider the simplifying limit $(\Omega_s, G_s)=(\infty, 0)$,
which is an endpoint of the instability line.
The interaction free energy 
in this limit
reads
\eq
F &=& \f{k_BT}{2\pi}\int_0^\infty q_{\perp} dq_{\perp} 
\biggl[
\ln\lrF{ 2 \sin \sqrt{q_0^2-q_\perp^2} d} \theta(q_0-q_\perp)
\switchmulticol{\nn\\&&\hspace{13mm}}
+\ln\lrS{1-e^{-2\sqrt{q_\perp^2-q_0^2}d}} \theta(q_\perp-q_0)
\biggr].
\label{Fsing}
\qe
Although it is a continuous function of $d$,
its gradient (and so $\Pi(d)$)
logarithmically diverges at $d=n\pi/q_0$ $(>0)$,
due to the contribution of the mode with
$(q_\perp,q_z) = (0, q_0)$.
This is closely related to the fact that
the mode $\phi \propto \sin(q_0 z)$,
which is soft in infinite bulk,
is allowed by the boundary condition $\phi(0)=\phi(d)=0$
only when $d=n\pi/q_0$.
When we gradually increase $d$ to cross one of the node points,
the soft mode is abruptly allowed and then prohibited again,
which corresponds to the rise and drop of the disjoining pressure.
For physical (finite) values of $\Omega_s$,
boundary constraints are not strict and 
the oscillation in $\Pi(d)$ is not singular~\cite{Ziherl2}.
%

%%%%%%%%%%%%%%%%%%%%%%%%%
% Linear surface field.
%%%%%%%%%%%%%%%%%%%%%%%%%

Up to now, we have assumed that the walls are ideally neutral.
In general, the walls have a specific interaction
with the fluid,
which is most simply described by
the linear Hamiltonian
$\Delta {\cal H}_s = \int d\br \;
h_s \phi \; \biglrL{\delta(z)+\delta(z-d)}$.
The surface field $h_s$ induces a mean-field interaction
between the walls.
Let us compare its magnitude
with that of the Casimir interaction.
To be specific, we consider symmetric diblock copolymer melts
with $N = 10^3$, $a = 0.5$ nm, and $T = 400$ K,
for which the lamella thickness $\pi/q_0$  equals $11$ nm.
For $\Omega_s=G_s=h_s/(\epsilon q_0^3)=1.0$, $p_0/q_0=0.4$,
and $d=4.7\pi/q_0=50$ nm,
minimization of $\cH=\cH_b + \cH_s + \Delta \cH_s$ gives
the surface value of the order parameter
$\phi = 0.54$
and the mean-field pressure $\Pi_{\rm mf}= -11$ Pa~\cite{NB1}.
On the other hand, the Casimir pressure for the same
parameters and at the same distance is $\Pi_{\rm ca} = -69$ Pa. 
The ratio $\Pi_{\rm ca}/\Pi_{\rm mf}=6.3$
shows that fluctuation effect on 
the total structural force
is significant within several lamella thicknesses.
It should be mentioned that harmonic approximation
to the free energy is valid only for a weak surface field. 
It would be realized, for instance,
by use of substrates coated by a thin sheet of 
random copolymers~\cite{KWMLRGS}.
For a strong surface field, the order parameter
is saturated at the walls where there appear
wetting layers of the lamellar phase~\cite{MM}.
In that case, the above estimate should be 
applicable to the force between the surfaces of 
the wetting layers. 
%

%
%%%%%%%%%%%%%%%%%
% Discussion
%%%%%%%%%%%%%%%%%

To summarize,
we predict an unconventionally
strong thermal Casimir effect in structured fluids
above the isotropic-lamellar transition.
At the ideally continuous transition,
the force has a long-range tail proportional to
$1/d$, which is due to the two-dimensional
distribution of bulk soft modes in the $\bq$-space.
The qualitative result should commonly hold
for model systems in which the bulk structure factor 
diverges at a finite wavenumber~\cite{Ziherl}.
Although the force cannot be genuinely 
long-ranged in a real system,
one should be able to access the temperature
region where the force has a range of several
layer thicknesses, within which it is much stronger 
than the conventional Casimir force at criticality.
This implies that it also dominates other dispersion 
forces, such as the van der Waals interaction, which is 
often comparable to the critical Casimir force~\cite{ZPZ,ML,GC}.
We have also studied the role of surface potential.
Within a standard model of microemulsions,
the force profile shows an oscillation
near the surface-induced I-L transition.
The oscillation originates in resonance of a
soft mode in the slab geometry.
A measurement of structural force
near the sponge-lamellar transition
has reported an oscillatory force superimposed on
an attractive background~\cite{AKR}.
However, more systematic study with controlled
surface properties seems necessary to extract
the fluctuation part of the interaction.
I wish the present work to stimulate 
further experimental investigation.

%
% ACKNOWLEDGMENT
%

I thank K. Sekimoto and K. Sato for useful discussions.
This work is supported in part by the Grant-in-Aid for
Scientific Research from Japan Society for the Promotion
of Science.

%
% REFERENCES
%

\switchmulticol{\end{multicols}}
\switchmulticol{\begin{multicols}{2}}
%
% FIGURES
%
\begin{figure}[h]
\vspace{35mm}
\epsfxsize=240pt \epsffile{casimir1.fig1.eps}
\mycaption{
Profile of the Casimir force
in the strong anchoring limit
$\omega_s \to \infty$, $g_s \to \infty$.
(a) Log-log plot of the absolute value $|\Pi|$
at the critical point $p_0=0$.
(b) Scaled disjoining pressure 
above the transition temperature.
}
\end{figure}
\begin{figure}[h]
\vspace{-35mm}
\epsfxsize=240pt \epsffile{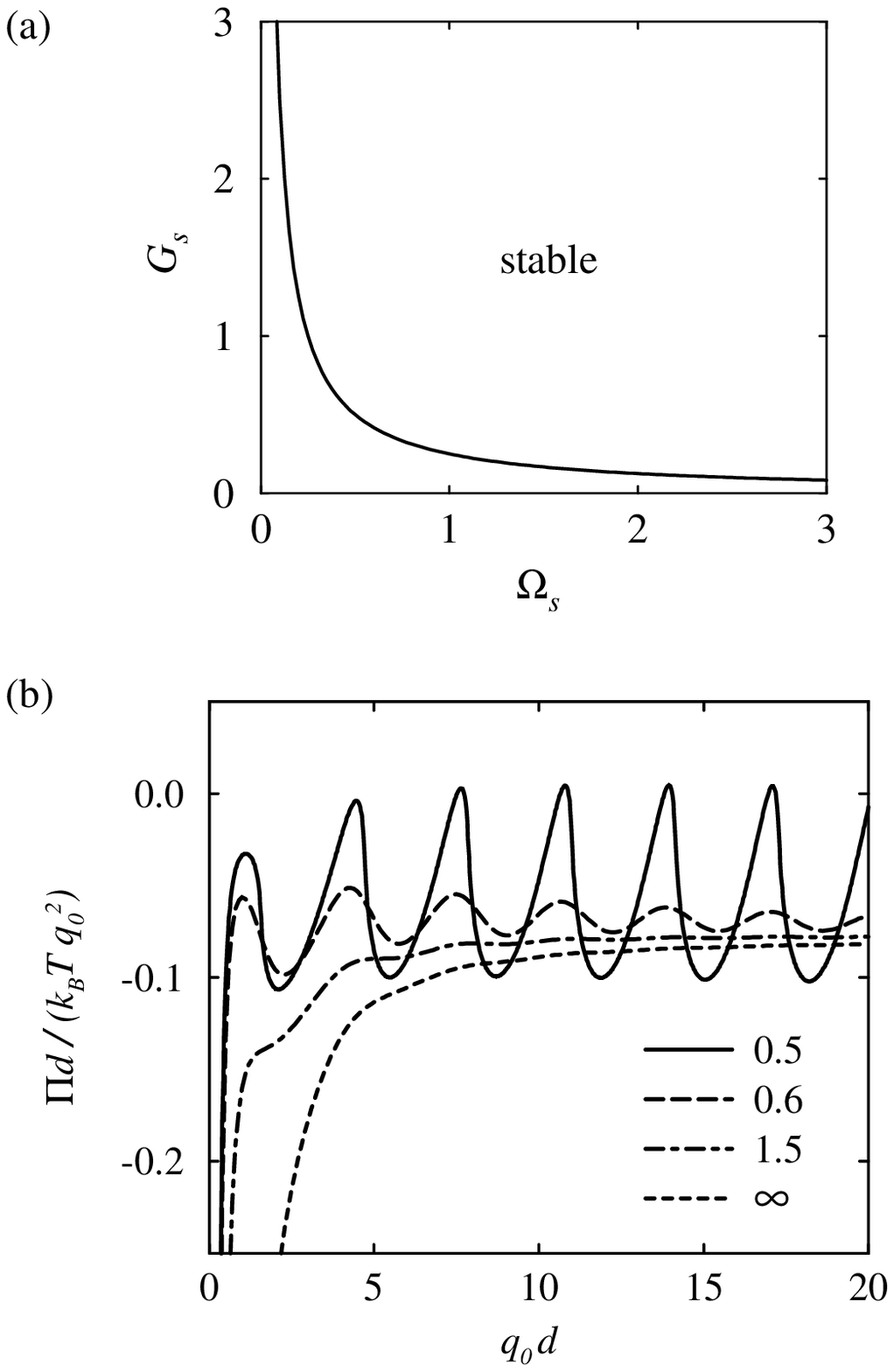}
\mycaption{
(a) Region of stability of the mean-field solution $\phi=0$.
Outside the region, the surface-induced I-L transition~[15]
preempts the bulk transition,
even if there is only one wall ($d\to\infty$).
(b) Surface-potential dependence of the Casimir pressure.
Plots are for
$(\Omega_s, G_s) = (0.5,0.5)$, $(0.6,0.6)$, $(1.5,1.5)$, and
$(\infty,\infty)$, all at the bulk transition temperature.
The point $(0.5,0.5)$ is on the instability line.
}
\end{figure}

\switchmulticol{\end{multicols}}

\begin{references}
\switchmulticol{\vspace{-15mm}}
\bibitem{Casimir}
H. B. G. Casimir, Proc. K. Ned. Akad. Wet. {\bf 51}, 793 (1948).

\bibitem{FdG}
M. E. Fisher and P. G. de Gennes, C. R. Acad. Sci. Ser. B 
{\bf 287}, 207 (1978).

\bibitem{Krech}
M. Krech, The Casimir Effect in Critical Systems (World Scientific, 
Singapore, 1994) and references therein.

\bibitem{Mikheev} L. V. Mikheev, Zh. \'Eksp. Teor. Fiz. {\bf 96},
632 (1989) [Sov. Phys. JETP {\bf 69}, 358 (1989)].

\bibitem{APP} A. Ajdari \etal, 
Phys. Rev. Lett. {\bf 66}, 1481 (1991); 
A. Ajdari \etal, 
J. Phys. II {\bf 2}, 487 (1992).

\bibitem{LK} 
H. Li and M. Kardar, 
Phys. Rev. Lett. {\bf 67}, 3275 (1991);
Phys. Rev. A {\bf 46}, 6490 (1992).

\bibitem{ZPZ}
P. Ziherl \etal, 
Phys. Rev. Lett. {\bf 82}, 1189 (1999);
{\it ibid} {\bf 84}, 1228 (2000). 

\bibitem{ML} A. Mukhopadhyay and B. M. Law,
Phys. Rev. Lett. {\bf 83}, 772 (1999).

\bibitem{GC}
R. Garcia and M. H. W. Chan, Phys. Rev. Lett. {\bf 83}, 1187 (1999).

\bibitem{VVC}
F. Vandenbrouck \etal, 
Phys. Rev. Lett. {\bf 82}, 2693 (1999);
see also Ref.[7] (latter).

\bibitem{KG}
M. Kardar and R. Golestanian, Rev. Mod. Phys. {\bf 71}, 1233 (1999).

\bibitem{Turner}
M. S. Turner, Phys. Rev. Lett. {\bf 69}, 1788 (1992).

\bibitem{SSKBSM}
M. Sikka \etal, 
Phys. Rev. Lett. {\bf 70}, 307 (1993). 

\bibitem{KWMLRGS}
G. J. Kellogg \etal, 
Phys. Rev. Lett. {\bf 76}, 2503 (1996). 

\bibitem{MM}
S. T. Milner and D. C. Morse, Phys. Rev. E {\bf 54}, 3793 (1996).

\bibitem{BFS}
K. Binder \etal, 
J. Phys. II {\bf 7}, 1353 (1997).

\bibitem{SS}
F. Schmid and M. Schick, Phys. Rev. E {\bf 48}, 1882 (1993).

\bibitem{HO} 
R. Ho{\l}yst and P. Oswald,
Phys. Rev. Lett. {\bf 79}, 1499 (1997).

\bibitem{CTM}
A. Ciach \etal, 
Europhys. Lett. {\bf 45}, 495 (1999).

\bibitem{Leibler}
L. Leibler, Macromolecules {\bf 13}, 1602 (1980).

\bibitem{TS}
M. Teubner and R. Strey, J. Chem. Phys. {\bf 87}, 3195 (1987).

\bibitem{ABJ}
D. Andelman \etal, 
J. Chem. Phys. {\bf 86}, 3673 (1987).

\bibitem{Swift}
J. Swift, Phys. Rev. A {\bf 14}, 2274 (1976).

\bibitem{GZ}
G. Gompper and S. Zschocke, Phys. Rev. A {\bf 46}, 4836 (1992).

\bibitem{Fierz}
M. Fierz, Helv. Phys. Acta {\bf 33}, 855 (1960).

\bibitem{Kleinert}
H. Kleinert, J. Math. Phys. {\bf 27}, 3003 (1986).

\bibitem{Brazovskii} S. A. Brazovski\v{i}, Zh. \'Eksp. Teor. Fiz. {\bf 68},
175 (1975) [Sov. Phys. JETP {\bf 41}, 85 (1975)];
G. H. Fredrickson and E. Helfand, J. Chem. Phys. {\bf 87}, 697 (1987). 


\bibitem{Ziherl2}
A similar logarithmic and repulsive singularity is 
found near a structural transition in nematic liquid crystals:
P. Ziherl \etal, Phys. Rev. E {\bf 61}, 5361 (2000).

\bibitem{NB1}
The mean-field force changes sign as a function of $d$, 
and this estimate is taken near a negative and local minimum.

\bibitem{Ziherl}
In contrast, for a presmectic fluid with monotonic 
correlation, the Casimir interaction is
shown to be of a conventional type:
P. Ziherl, Phys. Rev. E {\bf 61}, 4636 (2000).

\bibitem{AKR}
D. A. Antelmi \etal, 
J. Phys. II {\bf 5}, 103 (1995).


\end{references}
\end{document}